\documentclass[lettersize,journal]{IEEEtran}
\usepackage{amsmath,amsfonts}
\usepackage{algorithmic}
\usepackage{algorithm}
\usepackage{array}
\usepackage[caption=false,font=normalsize,labelfont=sf,textfont=sf]{subfig}
\usepackage{textcomp}
\usepackage{stfloats}
\usepackage{url}
\usepackage{verbatim}
\usepackage{graphicx}
\usepackage{cite}
\usepackage{lineno}
\usepackage{color}
\usepackage{ulem}
\usepackage{booktabs}
\usepackage{tabularx}
\usepackage[printonlyused,withpage]{acronym}
\acrodef{ldpc}[LDPC]{low-density parity-check}
\acrodef{snr}[SNR]{signal-to-noise ratio}
\acrodef{qam}[QAM]{quadrature amplitude modulation}
\acrodef{awgn}[AWGN]{additive white Gaussian noise}
\acrodef{ge}[GE]{Gilbert–Elliott}
\acrodef{bcjr}[BCJR]{Bahl–Cocke–Jelinek–Raviv}
\acrodef{ba}[BA]{burst-aware}
\acrodef{iba}[IBA]{iterative burst-aware}
\acrodef{dsp}[DSP]{digital signal processing}
\acrodef{eepn}[EEPN]{equalization-enhanced phase noise}
\acrodef{mpi}[MPI]{multipath interference}
\acrodef{sdfec}[SD-FEC]{soft-decision forward error correction}
\acrodef{per}[PER]{packet error rate}
\acrodef{llr}[LLR]{log-likelihood ratio}
\acrodef{ber}[BER]{bit error rate}
\acrodef{qpsk}[QPSK]{quadrature phase shift keying}
\acrodef{qam}[QAM]{quadrature amplitude modulation}
\acrodef{sova}[SOVA]{soft-output Viterbi algorithm}
\acrodef{bcjr}[BCJR]{Bahl–Cocke–Jelinek–Raviv}
\acrodef{va}[VA]{Viterbi algorithm}
\acrodef{mqam}[$M$-QAM]{$M$-ary quadrature amplitude modulation}
\acrodef{qpsk}[QPSK]{quadrature phase shift keying}
\begin{document}

\title{Channel Estimation and LDPC Decoding for Bursty Phase Noise}

\author{Han Cui, 
        Frank R. Kschischang,~\IEEEmembership{Fellow,~IEEE,} 
        Magnus Karlsson,~\IEEEmembership{Fellow,~IEEE,}
        Erik Agrell,~\IEEEmembership{Fellow,~IEEE} 
\thanks{This work was presented in part at the 2025 European Conference on Optical Communication [W.01.04.1](Corresponding author: Han Cui.)}
\thanks{This research was funded by the Swedish Research Council (VR) under grants no. 2021-03709.}
\thanks{Han Cui and Erik Agrell are with the Department of Electrical
Engineering, Chalmers University of Technology, 41296 Gothenburg, Sweden
(e-mail: hancu@chalmers.se; agrell@chalmers.se).}
\thanks{Frank R. Kschischang is with the Edward S. Rogers Sr. Department of Electrical and Computer Engineering, University of Toronto, Toronto, ON M5S 3G4, Canada (e-mail: frank@ece.utoronto.ca).}
\thanks{Magnus Karlsson is with the Department of Microtechnology and
Nanoscience, Chalmers University of Technology, 41296 Gothenburg, Sweden
(e-mail: magnus.karlsson@chalmers.se).}}



\maketitle

\begin{abstract}

Time-varying distortions in communication systems can significantly degrade the performance of soft-decision forward error correction. This paper presents a \ac{ba} \ac{ldpc} decoding scheme for channels affected by bursty phase noise. 
By applying differential coding to a Wiener process with time-varying innovation variance, bursty differential phase noise is obtained. Simulation results demonstrate that, compared to conventional decoding, the BA scheme achieves gains in the signal-to-noise ratio of up to $0.7$~dB at a \ac{ber} of $4\cdot10^{-3}$ and more than $1$~dB at a \ac{per} of $1\cdot10^{-2}$. Furthermore, by iterating between channel estimation and \ac{ldpc} decoding, forming the proposed \ac{iba} decoding scheme, the gains increase to $1.4$~dB and more than $3$~dB, respectively. More importantly, the IBA scheme significantly improves robustness to bursty phase noise. Compared with the conventional scheme, the IBA scheme can reduce both \ac{ber} and \ac{per} by up to two orders of magnitude under severe bursty phase noise.

\end{abstract}

\begin{IEEEkeywords}
Bursty channel, iterative receiver, \ac{ldpc} coding, phase noise.
\end{IEEEkeywords}
\acresetall
\section{Introduction}
\IEEEPARstart{I}{n} communication systems, impairments may vary over time due to physical effects, including component imperfections~\cite{clockburst,oscillator,laser} and environmental variations~\cite{envir,vibration1,vibration2,tem}, and also due to limitations in the \ac{dsp}~\cite{dspburst,dspburst2,dspburst3}.
These time-varying impairments may result in error bursts that can cause the performance of decoding algorithms to degrade~\cite{ldpcdecoding3}.

Phase noise is a common impairment in both wireless and optical communication systems. In wireless systems, it mainly originates from frequency instabilities of radio frequency oscillators \cite{wirelessphasenoise1}, while in optical systems, it is primarily caused by the finite linewidth of lasers \cite{opticalphasenoise1}. In both cases, phase noise can be accurately modeled as a Wiener process\cite{wirelessphasenoise2,opticalphasenoise2,opticalphasenoise3}, representing the cumulative nature of random frequency fluctuations over time. However, due to sudden changes in system components or imperfect phase-noise recovery, the phase noise may exhibit bursty characteristics in practical scenarios \cite{wirelessburst1,wirelessburst2,envir,laserburst,phaseburst}. Such bursts can be triggered by sudden changes in oscillator stability\cite{oscillator,laser}, mechanical vibrations\cite{vibration1,vibration2}, temperature fluctuations\cite{tem}, or even lightning strikes\cite{envir}. 

Besides phase noise, other types of bursty distortions can also occur in transmission systems.
In wireless fading channel systems, the channel is often modeled as a finite-state Markov channel because it exhibits temporal correlation and memory \cite{fading1,fading2}. In optical systems, phenomena such as polarization-mode dispersion, \ac{eepn}, and \ac{mpi} can lead to clusters of symbol errors~\cite{poloburst,sopburst,eepn1,eepn2,eepn3,mpiburst,mpi2}. Although these effects originate from different physical sources, they all lead to successive errors that are particularly detrimental to soft-decision forward error correction (SD-FEC) decoding\cite{eepn3,sdfec}.

\Ac{ldpc} codes are widely used for SD-FEC in modern communication systems because of their strong error-correcting capability and near-capacity performance~\cite{ldpc}. However, traditional \ac{ldpc} decoders are typically designed for \ac{awgn}, with no attempt made to accurately capture the temporal correlation of burst errors, leading to performance degradation. Moreover, decoding errors caused by burst errors are particularly critical because most network protocols perform error detection and retransmission at the packet level~\cite{plr}. Even a small number of concentrated bit errors can lead to the loss of entire packets, significantly increasing the \ac{per}. For bursty channels, a low-complexity approximate density evolution scheme was proposed in~\cite{ldpcdecoding1}, which performs channel estimation within the \ac{ldpc} decoding. Furthermore, a theoretical partial order relation for finite-state Markov channels was proposed in~\cite{ldpcdecoding2}. This theoretical framework clarifies how the memory of the Markov channel affects the \ac{ldpc} decoding performance.

Unlike previous studies that modify the internal structure of the decoding algorithm \cite{ldpcdecoding1,ldpcdecoding2,ldpcdecoding3,ldpcdecodinginside}, this paper employs an independent channel estimator before decoding. The estimated channel state information is incorporated into the \acp{llr} before being input into the \ac{ldpc} decoder. By refining the \acp{llr} to more accurately reflect the reliability of the received bits under different channel states, this method improves transmission performance in bursty channels while keeping the \ac{ldpc} decoding structure unchanged.

In~\cite{ecochan} we proposed a \ac{ba} \ac{ldpc} decoding scheme that integrates channel state estimation with \ac{ba} LLR calculation. This paper extends~\cite{ecochan} by extending the channel state estimation from hard-decision to soft-output processing. Furthermore, we extend the channel state estimation from hard-decision to soft-output processing. Furthermore, an \ac{iba} \ac{ldpc} decoding scheme is developed, which refines both the symbol and state probabilities through iterative information exchange between the channel estimator and the \ac{ldpc} decoder. 
The main contributions of this paper are summarized as follows:

\begin{itemize}
    \item \textbf{Channel modeling:} 
    We model the bursty phase noise channel as a \ac{ge} Markov-modulated Wiener process, where the Wiener process governs the continuous phase evolution, and the GE model introduces bursts in the phase noise~\cite{ge1}. 
After differential coding, the differential phase noise is well approximated as a zero-mean Gaussian process whose variance depends on the current GE state (either "good" or "bad").
    
    \item \textbf{Burst-aware decoding:} 
We propose a BA \ac{ldpc} decoding scheme that integrates channel state estimation with BA LLR computation. In this scheme, three channel estimation methods are considered, including the \ac{va}, the \ac{sova}, and the  \ac{bcjr} algorithm. Each method provides a different level of reliability information, which is then used to improve the accuracy of \ac{ldpc} decoding.

    \item \textbf{Iterative burst-aware decoding:} 
    Building upon the \ac{ba} \ac{ldpc} scheme, we further design an \ac{iba} \ac{ldpc} decoding structure that establishes a feedback loop between the channel estimator and the \ac{ldpc} decoder. 
    In each iteration, symbol probabilities are calculated from the decoder output LLRs, which can be used to improve channel state estimation, and the updated channel state probabilities are fed back to improve the accuracy of LLRs.
    As a result, the proposed \ac{iba} decoding scheme achieves significantly improved robustness and performance under severe bursty conditions.

    \item \textbf{Performance evaluation:} 
    Comprehensive simulations are conducted to compare the performance of conventional \ac{ldpc}, \ac{ba} \ac{ldpc}, and \ac{iba} \ac{ldpc} decoding schemes. 
    The analysis covers various dimensions, including modulation formats, \ac{snr} levels, burst severity, and average burst duration, demonstrating significant gains in both \ac{ber} and \ac{per} performance.
\end{itemize}

The remainder of this paper is organized as follows. 
Section~\ref{sec:system_model} introduces the system and channel models. 
Section~\ref{sec:ba_LDPC_decoding} presents the channel estimation schemes. 
Section~\ref{sec:interation} describes the \ac{ba} and \ac{iba} \ac{ldpc} decoding schemes. 
Simulation results are discussed in Section~\ref{sec:simulation}, and conclusions are drawn in Section~\ref{sec:conclusion}.
\section{System Model}
\label{sec:system_model}
\subsection{Bursty channel model}
\begin{figure}[t]
\centering
\includegraphics[scale=0.667]{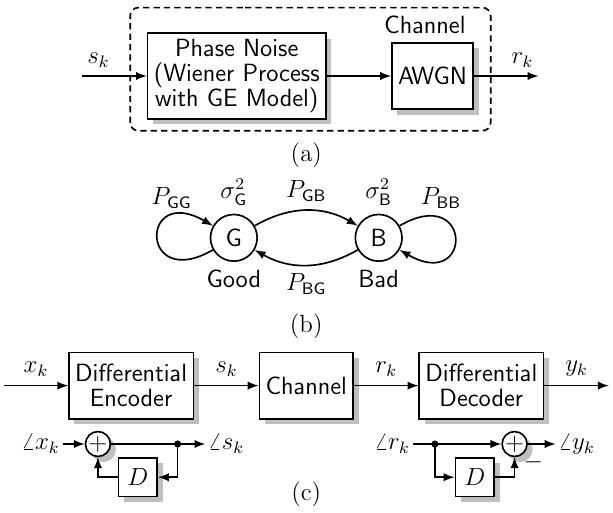}
\caption{Block diagram of the bursty differential phase noise channel model.(a) Bursty phase noise channel; (b) Gilbert–Elliott two-state Markov model; (c) phase-domain differential encoding and decoding.}
\label{fig:channel}
\end{figure}

The bursty phase noise channel model is illustrated in Fig.~\ref{fig:channel}(a). Let $s_k$ and $r_k$ denote the input and output of the channel, respectively. The received signal $r_k$ is affected by both AWGN and bursty phase noise, and can be written as
\begin{equation}
\label{eq:rx_model}
r_k = s_k e^{j\theta_k} + n_k,
\end{equation}
where $\theta_k$ denotes the phase noise and $n_k \sim \mathcal{CN}(0,\sigma^2)$ is complex circularly-symmetric Gaussian noise with zero mean and variance $\sigma^2$. Thus the real and imaginary parts of $n_k$ are independent zero mean Gaussian random variable, each having variance $\sigma^2/2$.

The phase noise $\theta_k$ is modeled as a Wiener process with parameters modulated by a two-state GE process.
The Wiener process captures the cumulative random walk behavior of oscillator phase drift\cite{wiener}, and is described as
\begin{equation}
\label{eq:wiener}
\theta_{k} = \theta_{k-1} + w_k, \quad w_k \sim \mathcal{N}(0, \sigma_{z_k}^2),
\end{equation}
where $w_k$ is the difference between two continuous phase noises values and $\sigma_{z_k}^2$ is its variance determined by the channel state $z_k$.  

To model the bursty characteristics, the GE process introduces two states $z_k \in \{\text{G},\text{B}\}$ as shown in Fig.~\ref{fig:channel}(b), representing the good and bad states, respectively\cite{ge1}. The state evolution follows a first-order Markov chain with transition probabilities

\begin{equation}
\label{eq:GE}
P(z_k = \text{B} | z_{k-1} = \text{G}) = P_{\text{GB}},
\end{equation}
\begin{equation}
\label{eq:GE}
P(z_k = \text{G} | z_{k-1} = \text{B}) = P_{\text{BG}},
\end{equation}
 and complementary probabilities $1-P_{\text{GB}}$ and $1-P_{\text{BG}}$ for remaining in the same state. The transition probabilities $P_{\text{GB}}$ and $P_{\text{BG}}$ control the expected number of consecutive time slots in which the Markov chain remains in the same state.  
The average number of symbols staying in each state can be expressed as
\begin{equation}
L_\text{G} = \frac{1}{P_{\text{GB}}}, \qquad
L_\text{B} = \frac{1}{P_{\text{BG}}},
\end{equation}
representing the mean durations of the good and bad states, respectively. 
The steady-state probabilities of being in the good or bad state 
can be calculated as
\begin{equation}
P_\text{G} = \frac{P_{\text{BG}}}{P_{\text{BG}}+P_{\text{GB}}}, \qquad
P_\text{B} = \frac{P_{\text{GB}}}{P_{\text{BG}}+P_{\text{GB}}},
\end{equation}
which represent the long-term fraction of time that the system spends in each state.

The variance of the innovation $w_k$ in the Wiener process~\eqref{eq:wiener} is then conditioned on $z_k$ as
\begin{equation}
\label{eq:var_state}
\sigma_{z_k}^2 =
\begin{cases}
\sigma_\text{G}^2, & z_k = \text{G}, \\
\sigma_\text{B}^2, & z_k = \text{B},
\end{cases}
\end{equation}
with $\sigma_\text{B}^2 \gg \sigma_\text{G}^2$, to reflect the significantly stronger differential phase noise experienced in the bad state.  

Overall, the mixed Wiener–GE model effectively characterizes both the slow random walk of phase drift and the occasional bursty fluctuations. 


\subsection{Differential coding}
In this work, differential encoding is performed only in the phase domain as shown in Fig.~\ref{fig:channel}(c), so no carrier phase recovery algorithm is required~[\citen{proakis2008digital}, Sec.~4.5].
Let $x_k$ and $r_k$ denote the symbols appearing at the input of the differential encoder and decoder, respectively, and let $s_k$ and $y_k$ denote the corresponding output symbols.
Let $\mathcal{X}$ denote the modulation constellation, and $M$ denotes the constellation size. The differential encoding and decoding process can be expressed as
\begin{equation}
\label{eq:diff_enc}
s_k = x_k \, e^{j\angle s_{k-1}},
\end{equation}
\begin{equation}
\label{eq:diff_dec}
y_k = r_k \, e^{-j\angle r_{k-1}}.
\end{equation}

According to the bursty channel model \eqref{eq:rx_model}, the phase of $r_{k}$ can be written as
\begin{equation}
\label{eq:rkphase}
\angle r_k = \left( \angle s_k + \theta_k +
\angle\!\left(1+\frac{n_k}{s_k}e^{-j\theta_k}\right)
\right)
\bmod 2\pi .
\end{equation}
From \eqref{eq:diff_dec} and \eqref{eq:rkphase}, the phase of $y_{k}$ can be expressed as 

\begin{align}
\angle y_k 
&= \angle r_k - \angle r_{k-1} 
\label{eq:11} \\
&= \angle s_k + \theta_k 
   + \angle\!\left(1+\frac{n_k}{s_k}e^{-j\theta_k}\right)
\notag  \\
&\quad
   - \angle s_{k-1} - \theta_{k-1}
   - \angle\!\left(1+\frac{n_{k-1}}{s_{k-1}}e^{-j\theta_{k-1}}\right)
 \label{eq:12}  \\
&= \biggl(\angle x_k + w_k
  + \angle\!\left(1+\frac{n_k}{s_k}e^{-j\theta_k}\right)  \notag \\
&\quad  
  - \angle\!\left(1+\frac{n_{k-1}}{s_{k-1}}e^{-j\theta_{k-1}}\right)\biggr)\bmod 2\pi  \label{eq:13},
\end{align}
where \eqref{eq:13} follows from \eqref{eq:wiener} and \eqref{eq:diff_enc}.
Then, the output of the differential decoding $y_k$ in \eqref{eq:diff_dec} can be calculated as

\begin{align}
y_k 
&= \bigl|s_k e^{j\theta_k}+n_k\bigr|\,e^{j\angle y_k} \label{eq:14} \\
&= |s_k|\,\left|1+\frac{n_k}{s_k}e^{-j\theta_k}\right|\,
   e^{j(\angle x_k+w_k)}
   e^{j\angle\left(1+\frac{n_k}{s_k}e^{-j\theta_k}\right)} \notag \\
&\quad 
   \cdot
   \exp\left(-j\angle\!\left(1+\frac{n_{k-1}}{s_{k-1}}e^{-j\theta_{k-1}}\right)\right) \label{eq:15}\\
&= x_k\,e^{j w_k}\left(1+\frac{n_k}{s_k}\,e^{-j\theta_k}\right) \notag \\
&\quad 
   \cdot   
   \exp\left(-j\angle\!\left(1+\frac{n_{k-1}}{s_{k-1}}e^{-j\theta_{k-1}}\right)\right) \label{eq:16} \\
&= \left(x_k\,e^{j w_k}
   + n_k e^{-j\left(\angle{s_{k-1}}+\theta_{k-1}\right)}\right) \notag \\
&\quad 
   \cdot
   \exp\left(-j\angle\!\left(1+\frac{n_{k-1}}{s_{k-1}}e^{-j\theta_{k-1}}\right)\right)\label{eq:17},
\end{align}
where we used \eqref{eq:diff_enc} to obtain \eqref{eq:16} and \eqref{eq:17}.
The first factor in \eqref{eq:17} can be interpreted as the transmitted signal affected by zero-mean Gaussian differential phase noise and AWGN. The second term $\exp\left(-j\angle\!\left(1+\frac{n_{k-1}}{s_{k-1}}e^{-j\theta_{k-1}}\right)\right)$ is an unpredictable phase rotation applied to the first term.

\begin{figure}
 \centering
 \includegraphics[scale=0.667]{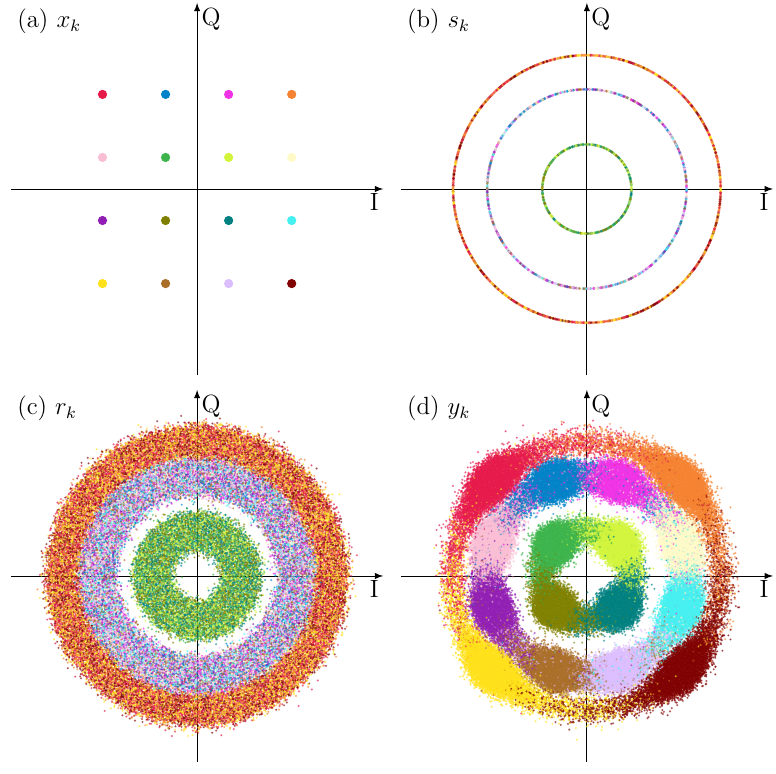}
 \caption{16QAM constellations for a fixed channel with $\sigma_\text{G}^{2} = 3 \cdot 10^{-4}$, $\sigma_\text{B}^{2} = 0.12$, 
 $P_\text{GB} = 2 \cdot 10^{-4}$, $P_\text{BG} = 2 \cdot 10^{-2}$, and \ac{snr} $= 20$~dB. (a)~Source signal $x_k$; (b)~signal after differential encoding $s_k$; (c)~received signal $r_k$; (d)~received signal after differential decoding $y_k$.}
 \label{fig:Constellation_16qam}
 \end{figure}

The constellation diagrams for 16QAM are presented in Fig.~\ref{fig:Constellation_16qam}.
Fig.~\ref{fig:Constellation_16qam} (a) and (b) display the transmitted symbols before and after differential encoding, respectively. After differential encoding, additional phases emerge due to the phase addition operation in the differential encoding, resulting in constellation diagrams that appear circular. Fig.~\ref{fig:Constellation_16qam}(c) displays the received symbols affected by bursty phase noise generated by the Wiener-GE model and AWGN. Fig.~\ref{fig:Constellation_16qam}(d) presents the symbols after differential decoding, which remain distorted due to differential phase noise and AWGN.

\begin{figure}
\centering
\includegraphics[scale=0.667]{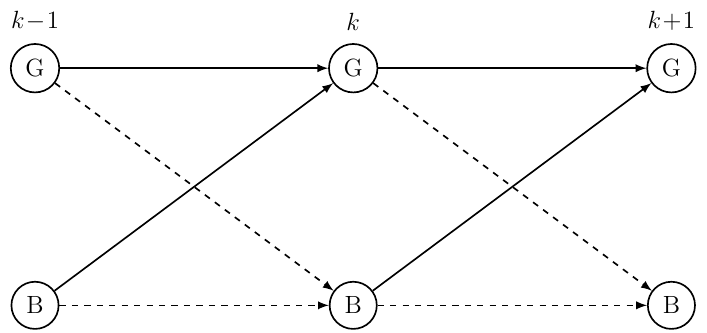}
\caption{Two-state trellis for channel state estimation algorithm.}
\label{fig:trellis}
\end{figure}
\section{Channel estimation}
\label{sec:ba_LDPC_decoding}

\begin{figure*}
\centering
\includegraphics[scale=0.667]{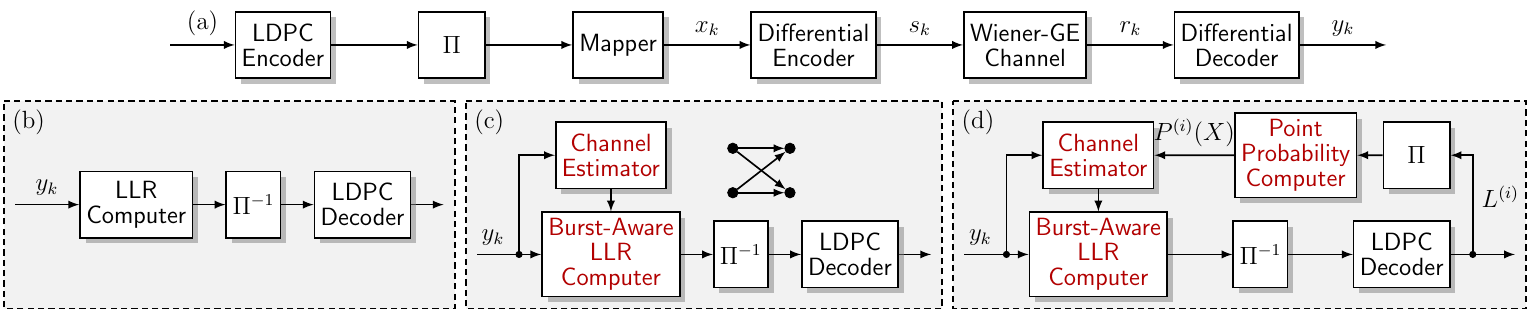}
\caption{(a) Schematic diagram of $M$-QAM transmission system with interleaving ($\Pi$), (b) baseline, (c) \ac{ba}, and (d) \ac{iba} \ac{ldpc} decoding.}
\label{fig:qam1}
\end{figure*}
To effectively mitigate the impact of bursty differential phase noise on \ac{ldpc} decoding performance, we estimate the channel state to obtain the sequence $z_k$. The goal of channel estimation is therefore to infer $z_k$ from the differentially decoded observation sequence $y_k$. Since the system model developed in Sec.~\ref{sec:system_model} does not yield a closed-form expression for the probability distribution of $y_k$, we adopt a simplified channel model to approximate the real channel during channel estimation. This model is expressed as
\begin{equation}
\label{eq:diff_ch}
y_k \approx x_k e^{j w_k} + \tilde{n}_k,
\qquad \tilde{n}_k \sim \mathcal{CN}(0,\tilde{\sigma}^2),
\end{equation}
where $\tilde{n}_k$ denotes the effective AWGN after differential encoding. We introduce a bias $\delta$ such that the effective noise variance satisfies $\tilde{\sigma}^2 = \delta \sigma^2$. The bias factor is an internal parameter in our algorithm that will be optimised later.

Since the exact probability density of $y_k$ in \eqref{eq:diff_ch} given $x_k$ and $z_k$ is analytically intractable, the authors of~\cite{llf} have derived two approximate expressions based on linear and bilinear transformations of the channel output. As shown in~\cite{llf}, the bilinear transformation (BLT) approximation provides better accuracy than the linear transform (LT) approximation. Our comparison in~\cite{ecochan} of the two approximations shows that the same is true for bursty channels. Therefore, the BLT-based approximation is adopted in this work, and it is given as
\begin{equation}
\label{eq:BLT}
\begin{aligned}
\log p(y_k|x_k,z_k)
\approx 
&-\frac{1}{\tilde{\sigma}^2}|y_k-x_k|^2\\
&+\frac{4\sigma_{z_k}^2}{2\tilde{\sigma}^4 + \tilde{\sigma}^2\sigma_{z_k}^2|x_k+y_k|^2}
\big(\Im\{x_k^{*}y_k\}\big)^2 \\
&-\frac{1}{2}\log\!\big(\tilde{\sigma}^2+\frac{\sigma_{z_k}^2}{2}|x_k+y_k|^2\big) + \log \left( \pi \tilde{\sigma}\right).
\end{aligned}
\end{equation}

We study three channel estimation schemes based on the state-dependent likelihood model: 1) \ac{va}~\cite{VA}, 2) \ac{sova}~\cite{SOVA}, and 3) \ac{bcjr}~\cite{BCJR}. 

\subsubsection{\ac{va}}
The \ac{va} scheme employs a two-state trellis, shown in Fig.~\ref{fig:trellis}, corresponding to the GE model. Each branch represents a different transition between two channel states. Branch metrics are computed as the negative log-likelihood of the corresponding state transition
\begin{equation}
\label{eq:hd_metric_marginal}
\gamma_k(z_{k-1}\!\to\!z_k)
= -\log p\!\left(y_k \middle| z_k\right)
- \log P\!\left(z_k \middle| z_{k-1}\right),
\end{equation}
where the likelihood $p\!\left(y_k \middle| z_k\right)$ is obtained
by summing over all possible transmitted symbols as
\begin{equation}
\label{eq:hd_marginal}
p\!\left(y_k \middle| z_k\right)
= \sum_{x_k\in\mathcal{X}}
P\!\left(x_k\right)\,
p\!\left(y_k \middle| x_k, z_k\right).
\end{equation}
The symbol probability $P\!\left(x_k\right)$ is assumed to be uniform over the constellation, thus $P\!\left(x_k\right) = 1/M$. This marginalization allows the VA to calculate the state transition likelihoods without knowledge of the transmitted symbols, and $p(y_k|x_k,z_k)$ is obtained from \eqref{eq:BLT}.
The VA identifies the state sequence $\hat{\mathbf z}$ by
recursively minimizing the accumulated path metric,
\begin{equation}
\Gamma_k(z_k)
=
\min_{z_{k-1}\in\{G,B\}}
\left[
\Gamma_{k-1}(z_{k-1})
+
\gamma_k\!\left(z_{k-1}\rightarrow z_k\right)
\right].
\end{equation}

The estimated $\hat{z}_k$ indicates whether the channel is in a good or bad state at each time, from which the corresponding differential phase noise variance $\sigma_{z_k}^2$ can be determined. 
The ensuing LLR computation assumes that the state estimates provided by the VA are correct, setting $P(z_k= G) = 1$ if $\hat{z}_k = G$ and setting $P(z_k = B) = 1$ if $\hat{z}_k = B$.

\subsubsection{\ac{sova}}
 The \ac{sova} extends the VA by providing the reliability of each state decision. At each trellis stage, both the best path and the nearest competing path are tracked. The reliability for the state at time $k$ is expressed as
\begin{equation}
\label{eq:sova_llr}
\Delta(z_k) \approx 
\Gamma_k^{\text{alt}} - \Gamma_k^{\text{best}},
\end{equation}
where $\Gamma_k^{\text{best}}$ and $\Gamma_k^{\text{alt}}$ denote the accumulated metrics of the survivor and the closest alternative path differing in $z_k$, respectively. $\hat{z}_k$ and $\Delta(z_k)$ represent the most likely state and the confidence, respectively. The probability of the G state used in LLR calculation is given as
\begin{equation}
P(z_k=\text{G}) =
\begin{cases}
\frac{1}{1+\exp{(-\Delta(z_k)})}, & \hat{z}_k=\text{G}, \\
\frac{1}{1+\exp{(\Delta(z_k)})}, & \hat{z}_k=\text{B},
\end{cases}
\end{equation}
and the probability of the B state is equal to $1-P(z_k=\text{G})$.

\subsubsection{\ac{bcjr}}
The posteriori state probabilities can be computed using the \ac{bcjr} algorithm, which performs a forward-backward recursion on the two-state trellis. The forward and backward metrics are computed as
\begin{equation}
\label{eq:bcjr_alpha}
\alpha_k(z_k)=
\sum_{z_{k-1}}\alpha_{k-1}(z_{k-1})
\,P(z_k|z_{k-1})\,p(y_k|z_k),
\end{equation}
\begin{equation}
\label{eq:bcjr_beta}
\beta_{k-1}(z_{k-1})=
\sum_{z_k}\beta_k(z_k)
\,P(z_k|z_{k-1})\,p(y_k|z_k),
\end{equation}
leading to a posteriori probabilities that can be directly used for LLR calculation in the form 
\begin{equation}
P(z_k) = C_k \alpha_k(z_k) \beta_k(z_k),
\end{equation}
where $C_k$ is a normalizing constant that ensures that the state probabilities sum to one.

The three estimation schemes have different output characteristics and achievable performance. Their outputs are unified into probabilities of being in the G or B state before being fed into the LLR calculation. These probabilities serve as weights for the likelihood calculation. Section~\ref{sec:simulation} compares their performance under varying \ac{snr} scenarios.
\section{Decoding schemes}
\subsection{\Ac{ba} \ac{ldpc} decoding}
We propose the \ac{ba} \ac{ldpc} decoding scheme as shown in Fig.~\ref{fig:qam1}(c). Compared with the baseline transmission systems shown in Fig.~\ref{fig:qam1}(b), the proposed scheme consists of two components, which are a channel state estimation algorithm and a \ac{ba} LLR calculation.
After obtaining channel state probabilities $P\!\left(z_k\right)$ in the channel estimation, the \ac{ba} likelihood function can be calculated as
\begin{equation}
\label{eq:soft_likelihood}
p\!\left(y_k \middle| x_k\right)
=\!\!\sum_{z_k\in\{\text{G},\text{B}\}}\! 
P\!\left(z_k\right)\,
p\!\left(y_k \middle| x_k, z_k\right),
\end{equation}
 where the state-dependent likelihood $p(y_k|x_k,z_k)$ is taken from \eqref{eq:BLT}.
Then, the LLR for the $m$-th bit of a transmitted symbol can be expressed as
\begin{equation}
\label{eq:llr}
\text{LLR}_m(y_k)
= \log 
\frac{\displaystyle \sum_{x_k \in \mathcal{X}_m^{(0)}}
  p(y_k|x_k) P(x_k)}
{\displaystyle \sum_{x_k \in \mathcal{X}_m^{(1)}}
  p(y_k|x_k) P(x_k)},
\end{equation}
where $\mathcal{X}_m^{(0)}$ and $\mathcal{X}_m^{(1)}$ denote the sets of constellation symbols whose $m$-th bit is 0 or 1, respectively. Assuming uniformly distributed symbols, $P(x_k)=1/M$ can be omitted since it cancels in the ratio.

As a baseline, when the channel-state information is unavailable, 
the likelihood $p(y_k|x_k)$ can be calculated using $\sigma_{\text{eff}}^2$, 
which represents an effective variance

\begin{equation}
\label{eq:eff_var}
\sigma_\text{eff}^2 = P_\text{G}\sigma_\text{G}^2 + P_\text{B}\sigma_\text{B}^2,
\end{equation}
where $P_\text{G}$ and $P_\text{B}$ are the steady-state probabilities of the good and bad states, respectively. This formulation corresponds to a memoryless approximation, ignoring burst correlations.

\subsection{\Ac{iba} \ac{ldpc} decoding}
\label{sec:interation}

\begin{figure}
\centering
\includegraphics[scale=0.667]{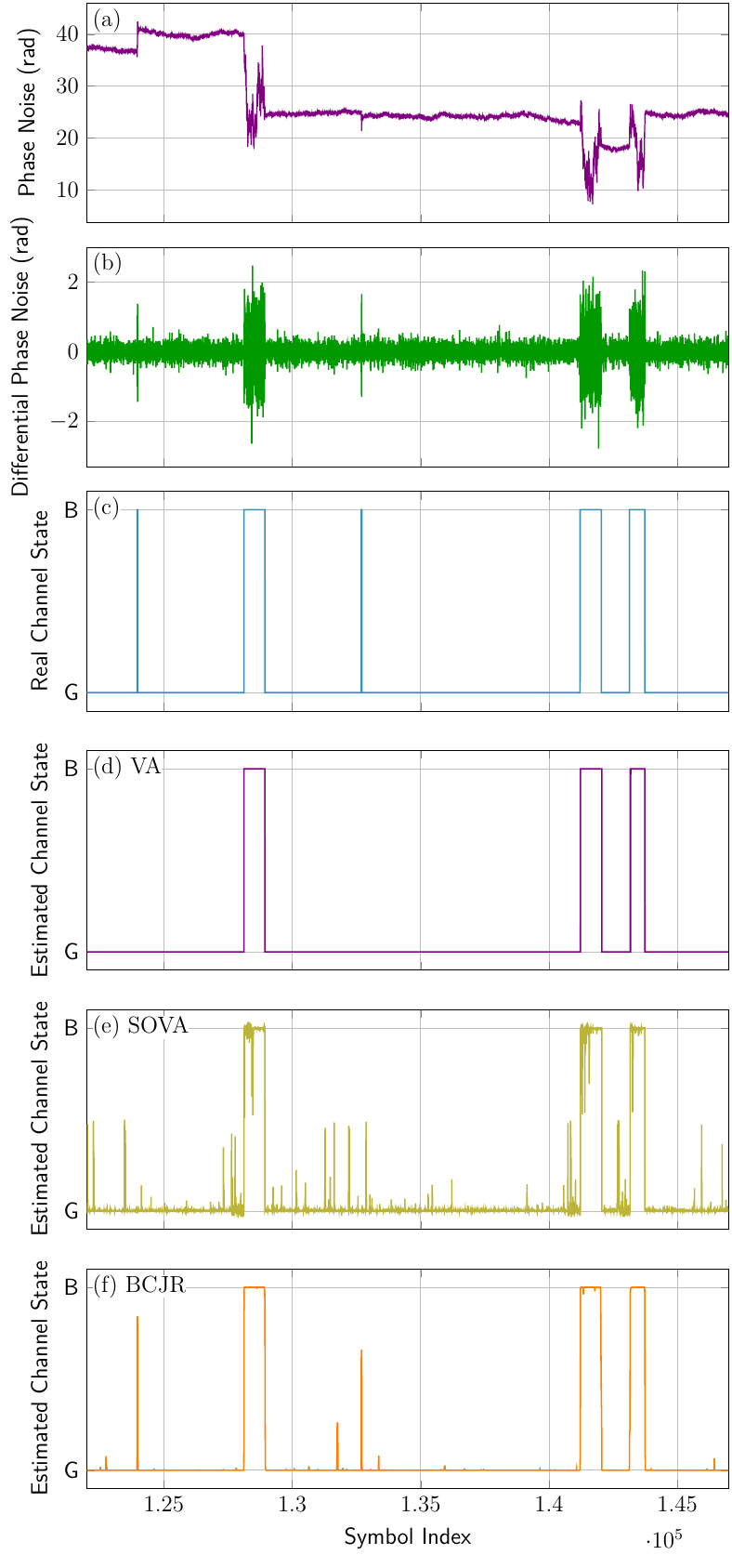}
\caption{Channel information (a)–(c) at \ac{snr} = $20$~dB and the estimated channel states (d)–(f) obtained at \ac{snr} = $15.5$~dB, for fixed parameters $\delta = -3$~dB, $\sigma_\text{G}^{2} = 3 \cdot 10^{-4}$, $\sigma_\text{B}^{2} = 0.12$, $P_\text{GB} = 2 \cdot 10^{-4}$, and $P_\text{BG} = 2 \cdot 10^{-3}$. (a)~Phase noise generated by the Wiener–GE model; (b)~differential phase noise after differential decoding; (c)~channel state generated by the GE model; (d)~estimated channel state using \ac{va}; (e)~estimated channel state using \ac{sova}; and (f)~estimated channel state using \ac{bcjr}.}
\label{fig:channel_information}
\end{figure}

To further improve the decoding performance in bursty differential phase-noise channels, an \ac{iba} receiver is developed based on the proposed \ac{ba} \ac{ldpc} decoding scheme, as illustrated in Fig.~\ref{fig:qam1}(d). The proposed scheme involves two forms of iteration. The first occurs between channel estimation and \ac{ldpc} decoding, whereas the second occurs internally within the \ac{ldpc} decoder. For clarity, we refer to the former as outer iterations and to the latter as decoding iterations. The \ac{ldpc} decoder is reinitialized at the beginning of each outer iteration.

In each outer iterations, the channel estimator produces state probabilities and symbol-wise likelihoods, from which \ac{ba} LLRs are computed and passed to the \ac{ldpc} decoder. 
The output LLRs of the decoder are then converted into symbol probabilities and fed back to the channel estimator for the next outer iteration. 
More accurate symbol probabilities not only improve the accuracy of the channel estimation, but also directly improve the accuracy of the LLRs calculation. Overall, the iterative process yields more accurate symbol and channel state probabilities, thus improving the robustness to bursty differential phase noise.

At the beginning stage of the \ac{iba} scheme, the channel estimator obtains the state probabilities $P\!\left(z_k\right)$, and the corresponding likelihood values $p\!\left(y_k \middle| x_k\right)$. Using this information, the \ac{ba} LLR calculation provides the input LLRs for the \ac{ldpc} decoder. 
The bias used during outer iterations can be denoted as $\delta'$ to match the channel, and the effective noise variance in \eqref{eq:BLT} is replaced by $\delta'\sigma^2$. 
After deinterleaving, these LLRs are fed into the \ac{ldpc} decoder for decoding, updating the decoded LLRs $L^{(i)}$ in the $i$th outer iteration.

\begin{figure*}
\centering
\includegraphics[scale=0.667]{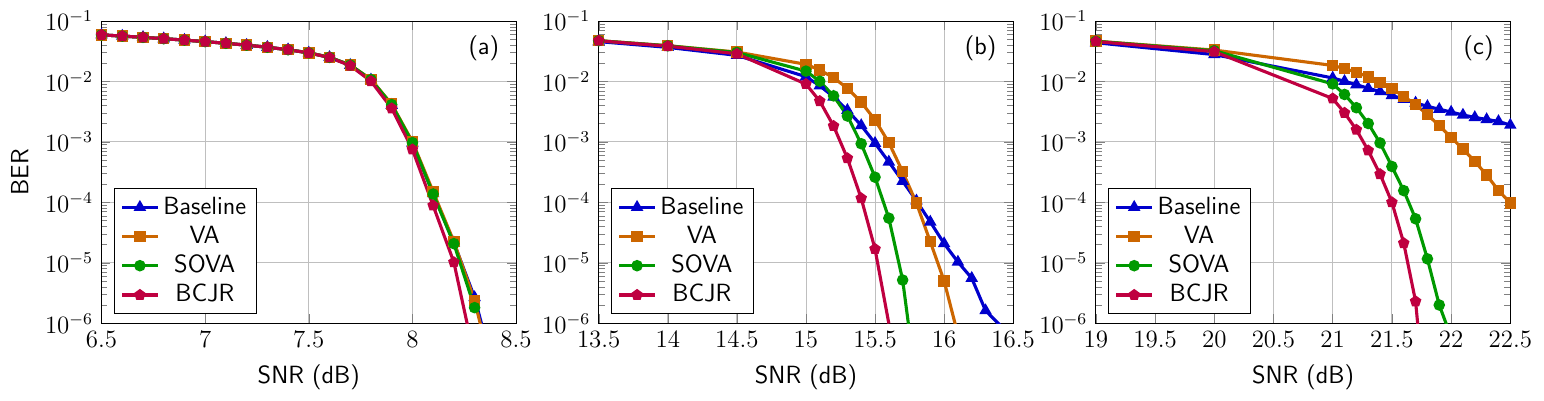}
\caption{BER performance of the baseline and three \ac{ba} \ac{ldpc} decoding schemes under different \acp{snr}, for a fixed $\sigma_\text{G}^{2} = 3 \cdot 10^{-4}$, $\sigma_\text{B}^{2} = 0.12$, $P_\text{GB} = 2 \cdot 10^{-4}$, and $P_\text{BG} = 2 \cdot 10^{-2}$. (a)~QPSK; (b)~16QAM; (c)~64QAM.}
\label{fig:snrber}
\end{figure*}

\begin{figure*}
\centering
\includegraphics[scale=0.667]{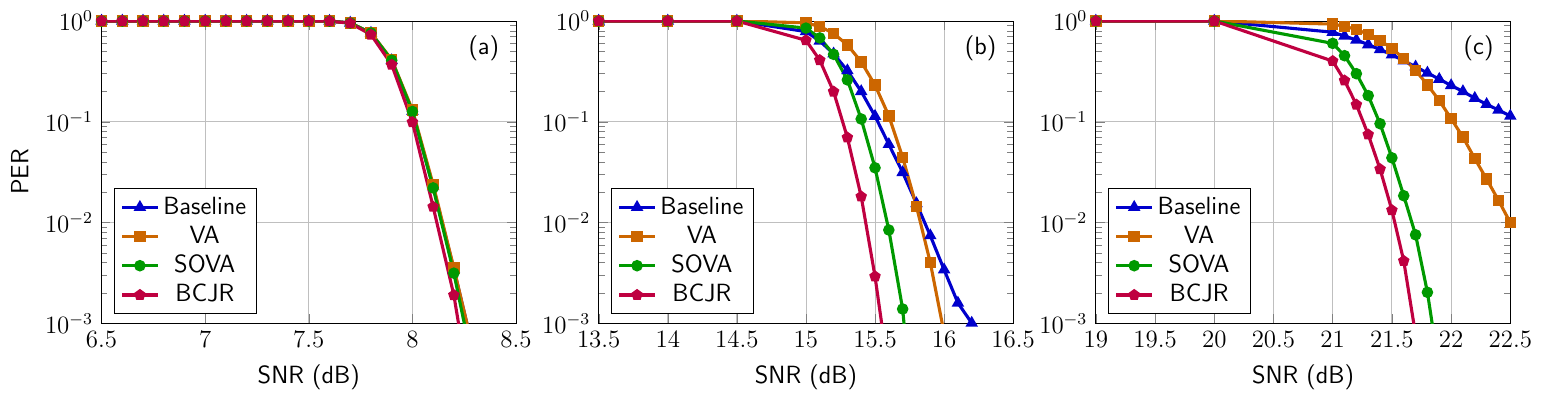}
\caption{PER performance of the baseline and three \ac{ba} \ac{ldpc} decoding schemes under different \acp{snr}, for a fixed $\sigma_\text{G}^{2} = 3 \cdot 10^{-4}$, $\sigma_\text{B}^{2} = 0.12$, $P_\text{GB} = 2 \cdot 10^{-4}$, and $P_\text{BG} = 2 \cdot 10^{-2}$. (a)~QPSK; (b)~16QAM; (c)~64QAM.}
\label{fig:snrper}
\end{figure*}

In the next outer iteration, the updated LLRs are re-interleaved and converted into symbol probabilities for channel estimation.
The bit probabilities at the $i$th outer iteration are given by
\begin{equation}
P\!\left(b^{(i)}_{k,m}=0\right)
=
\frac{1}{1+\exp\!\left(-L^{(i)}_{k,m}\right)},
\end{equation}
\begin{equation}
P\left(b^{(i)}_{k,m}=1\right)=1-P\left(b^{(i)}_{k,m}=0\right).
\end{equation}


Let $\mathcal{X}\!=\!\left\{c_1,\ldots,c_M\right\}$ be the constellation and 
$\mathbf{b}\!\left(c_j\right)\!=\!\left[b_1\!\left(c_j\right),\ldots,b_{m}\!\left(c_j\right)\right]$ 
is the bit label of $c_j$. Assuming independent bit probabilities, the probability for $x_k\!=\!c_j$ is
\begin{equation}
\label{eq:symbol_weight}
P^{(i)}_k\!\left(c_j\right)
=\prod_{m=1}^{\text{log}_2(M)} P\!\left(b^{(i)}_{k,m}=b_m\!\left(c_j\right)\right).
\end{equation}
The normalized symbol probability in the $i$th outer iteration is then
\begin{equation}
\label{eq:symbol_prior_from_bits}
P^{(i)}\!\left(x_k=c_j\right)
=\frac{P^{(i)}_k\!\left(c_j\right)}{\displaystyle\sum_{\ell=1}^{M} P^{(i)}_k\!\left(c_\ell\right)},
\end{equation}
which is then used in place of the uniform distribution $P(x_k)$ in the
channel-state estimation \eqref{eq:hd_marginal} and LLR calculation \eqref{eq:llr} of the $i{+}1$th outer iteration.


\section{Results}
\label{sec:simulation}
Monte Carlo simulations were conducted for \ac{ldpc}-coded $M$-QAM transmission over a mixed Wiener–GE channel affected by AWGN. The employed \ac{ldpc} code conforms to the IEEE 802.3ca standard, with a codeword length of 17664 bits and 14592 information bits\cite{802ldpc}. The number of decoding iterations is fixed at 15.
A total of 52531200 information bits were transmitted, corresponding to approximately 102600 packets, each containing 512 bits\cite{512packet}. 
The interleaver is implemented as a block interleaver constructed by reshaping the bits into 1024 rows and reading it out column-wise at the transmitter, with the inverse operation applied at the receiver. Three modulation formats were considered: \ac{qpsk} with $M=4$, and \ac{mqam} with $M=16$ and $M=64$.

The simulation results are presented in three parts. The first part analyzes the channel characteristics, including the evolution of phase noise, channel-state transitions, differential phase noise distribution, and the performance of three estimation algorithms (VA, SOVA, and BCJR). The second part evaluates the \ac{ba} \ac{ldpc} decoding schemes by comparing their BER and PER performance with the baseline decoder under different modulation formats, as well as by optimizing the \ac{snr} bias $\delta$. The third part focuses on the \ac{iba} decoding, in which the initial bias $\delta$ is the same as in BA, and another bias $\delta^{\prime}$ is used during the iterations. A comprehensive comparison among the baseline, \ac{ba}, and \ac{iba} \ac{ldpc} decoders is then performed under five different channel dimensions by varying key parameters of the bursty differential phase noise channel, including the \ac{snr}, the phase-noise variances $\sigma_\text{B}^2$ and $\sigma_\text{G}^2$, and the state transition probabilities $P_\text{GB}$ and $P_\text{BG}$. 

\begin{figure}
\centering
\includegraphics[scale=0.667]{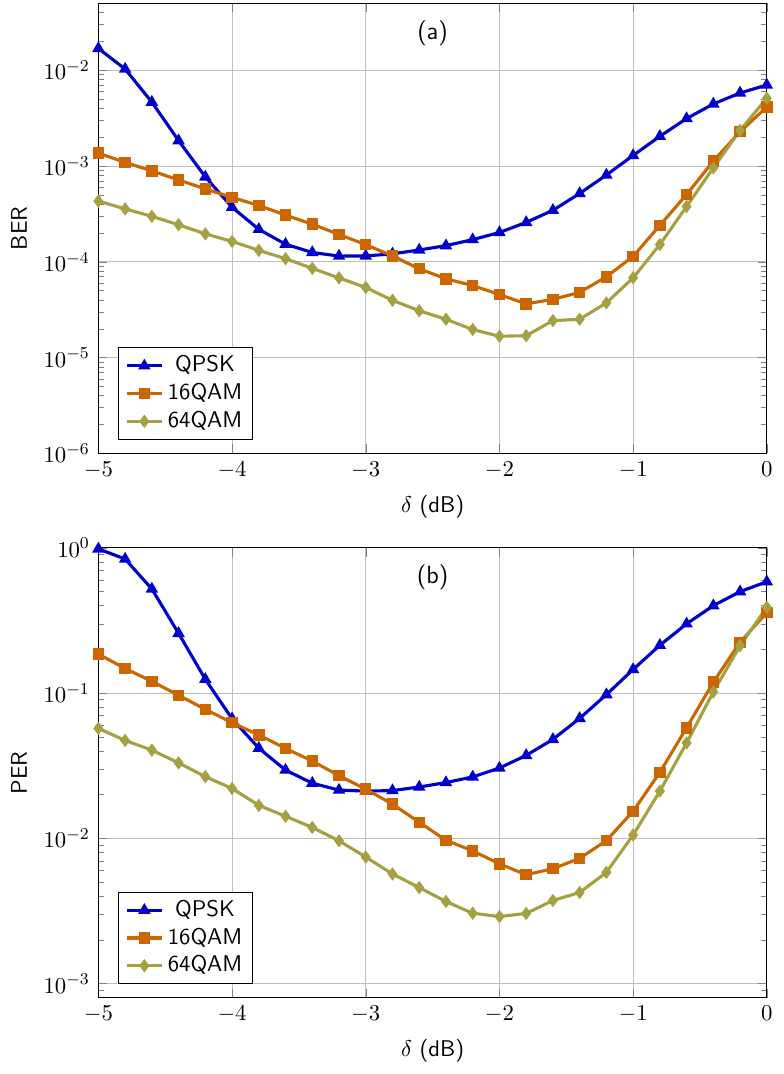}
\caption{Bias $\delta$ optimization under different modulation formats, for a fixed $\sigma_\text{G}^{2} = 3 \cdot 10^{-4}$, $\sigma_\text{B}^{2} = 0.12$, $P_\text{GB} = 2 \cdot 10^{-4}$, and $P_\text{BG} = 2 \cdot 10^{-2}$. The \ac{snr} values are set to $8.1$~dB for QPSK, $15.4$~dB for 16QAM, and $21.5$~dB for 64QAM.}
\label{fig:Bias1opt}
\end{figure}
\subsection{System model}

\begin{figure*}
\centering
\includegraphics[scale=0.667]{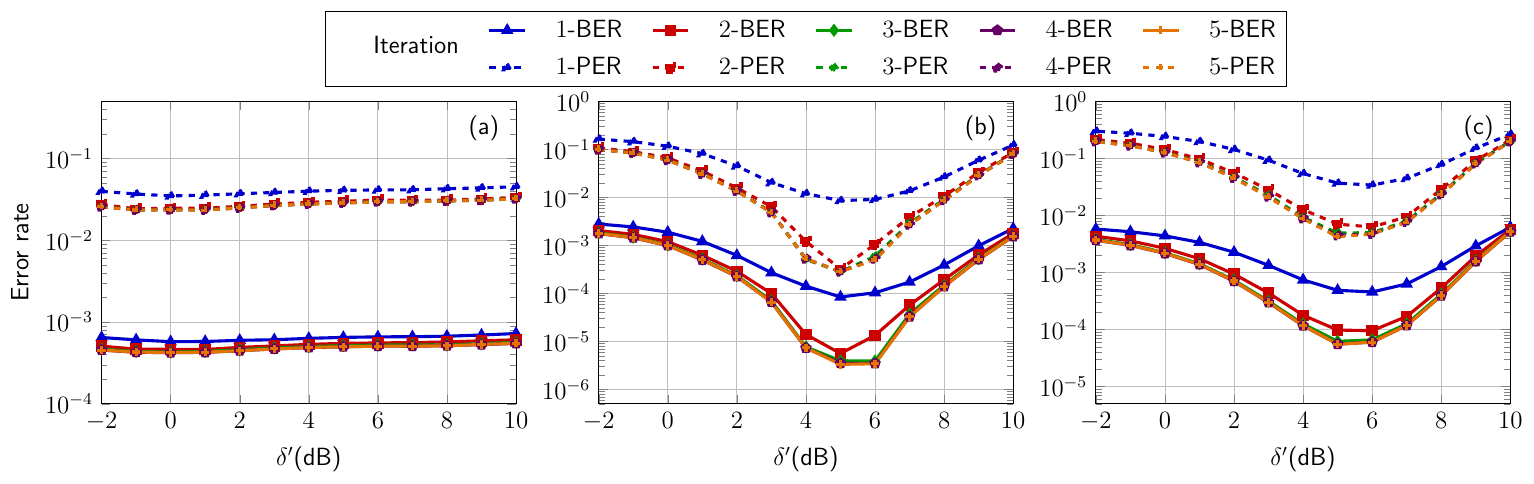}
\caption{Optimization of the bias $\delta^{\prime}$ parameter in the iterative decoding process for 
(a)~QPSK, (b)~16QAM, and (c)~64QAM. 
The simulations were conducted for a fixed set of channel parameters 
$\sigma_\text{G}^{2} = 3\cdot10^{-4}$, $\sigma_\text{B}^{2} = 0.12$, 
$P_\text{GB} = 2\cdot10^{-4}$, and $P_\text{BG} = 2\cdot10^{-2}$, 
with \acp{snr} of 8~dB, 15~dB, and 20.8~dB for QPSK, 16QAM, and 64QAM, respectively.}
\label{fig:bias2}
\end{figure*}

\begin{figure*}
\centering
\includegraphics[scale=0.667]{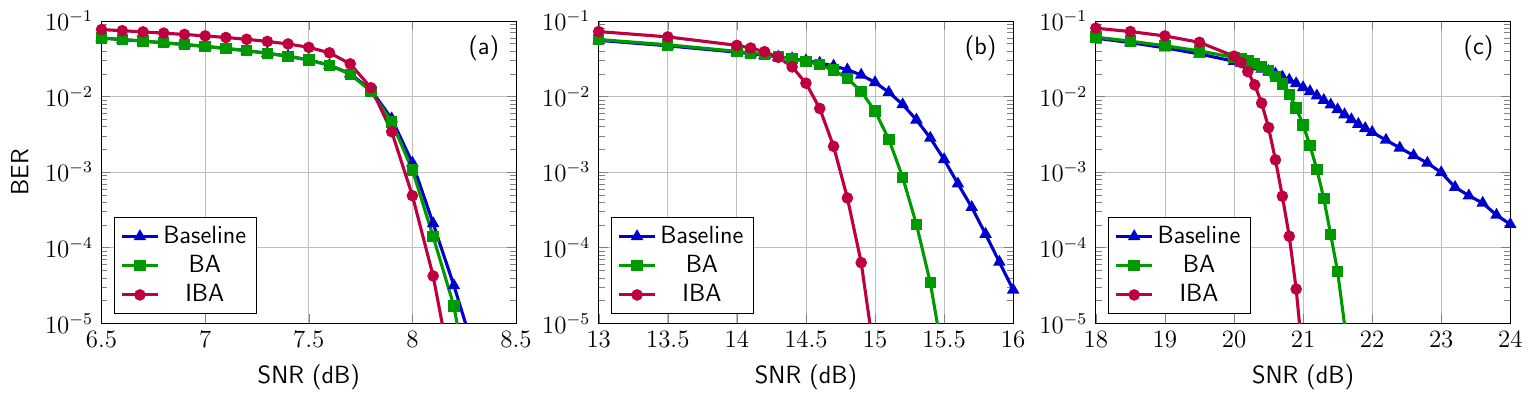}
\caption{BER performance of the baseline, \ac{ba} and \ac{iba} \ac{ldpc} decoding schemes under different \acp{snr}, for a fixed $\sigma_\text{G}^{2} = 3 \cdot 10^{-4}$, $\sigma_\text{B}^{2} = 0.12$, $P_\text{GB} = 2 \cdot 10^{-4}$, and $P_\text{BG} = 2 \cdot 10^{-2}$. (a)~QPSK; (b)~16QAM; (c)~64QAM.}
\label{fig:ibasnrber}
\end{figure*}

\begin{figure*}
\centering
\includegraphics[scale=0.667]{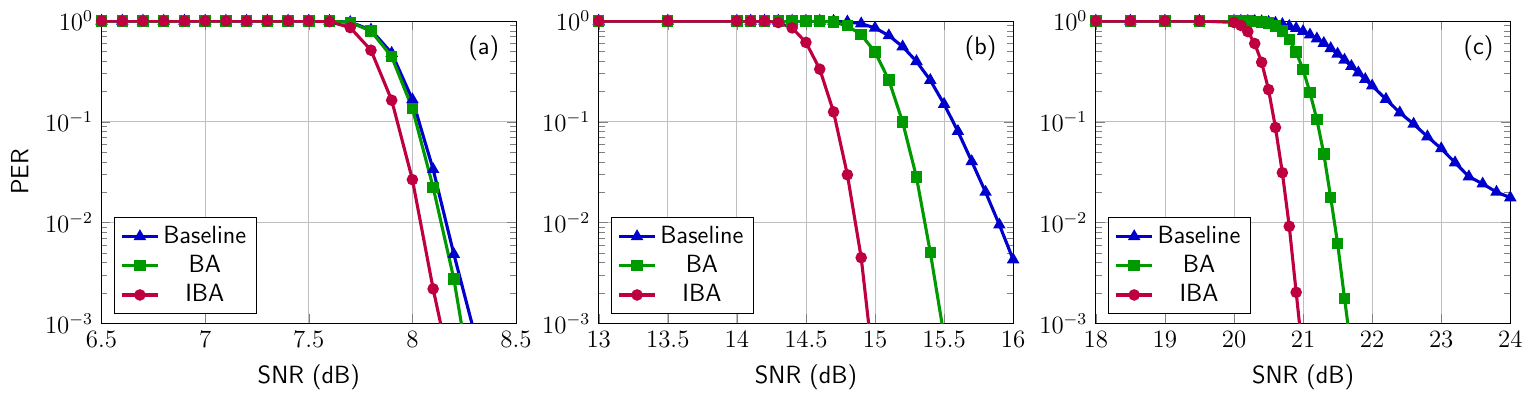}
\caption{PER performance of the baseline, \ac{ba} and \ac{iba} \ac{ldpc} decoding schemes under different \acp{snr}, for a fixed $\sigma_\text{G}^{2} = 3 \cdot 10^{-4}$, $\sigma_\text{B}^{2} = 0.12$, $P_\text{GB} = 2 \cdot 10^{-4}$, and $P_\text{BG} = 2 \cdot 10^{-2}$. (a)~QPSK; (b)~16QAM; (c)~64QAM.}
\label{fig:ibasnrper}
\end{figure*}

\begin{figure*}
\centering
\includegraphics[scale=0.667]{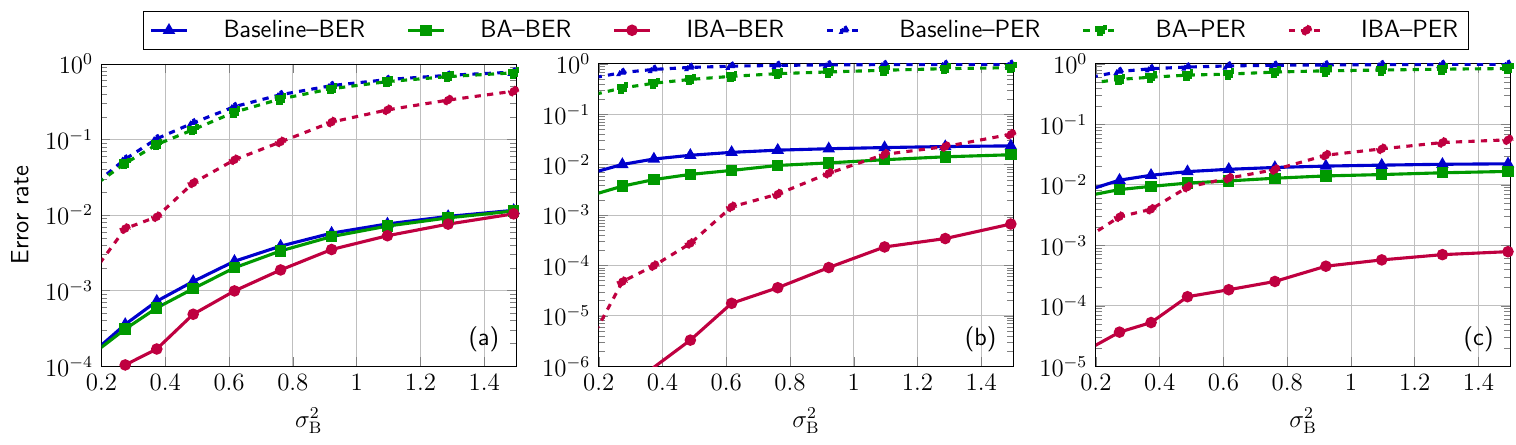}
\caption{BER and PER performance of the baseline, \ac{ba}, and \ac{iba} \ac{ldpc} decoding schemes under different $\sigma_\text{B}^{2}$ values, with fixed parameters 
$\sigma_\text{G}^{2} = 3 \cdot 10^{-4}$, $P_\text{GB} = 2 \cdot 10^{-4}$, 
and $P_\text{BG} = 2 \cdot 10^{-2}$. 
(a)~QPSK with $\text{SNR}  = 8$~dB; 
(b)~16QAM with $\text{SNR}  = 15$~dB; 
(c)~64QAM with $\text{SNR} = 20.8$~dB.}
\label{fig:ibasigmab}
\end{figure*}

\begin{figure*}
\centering
\includegraphics[scale=0.667]{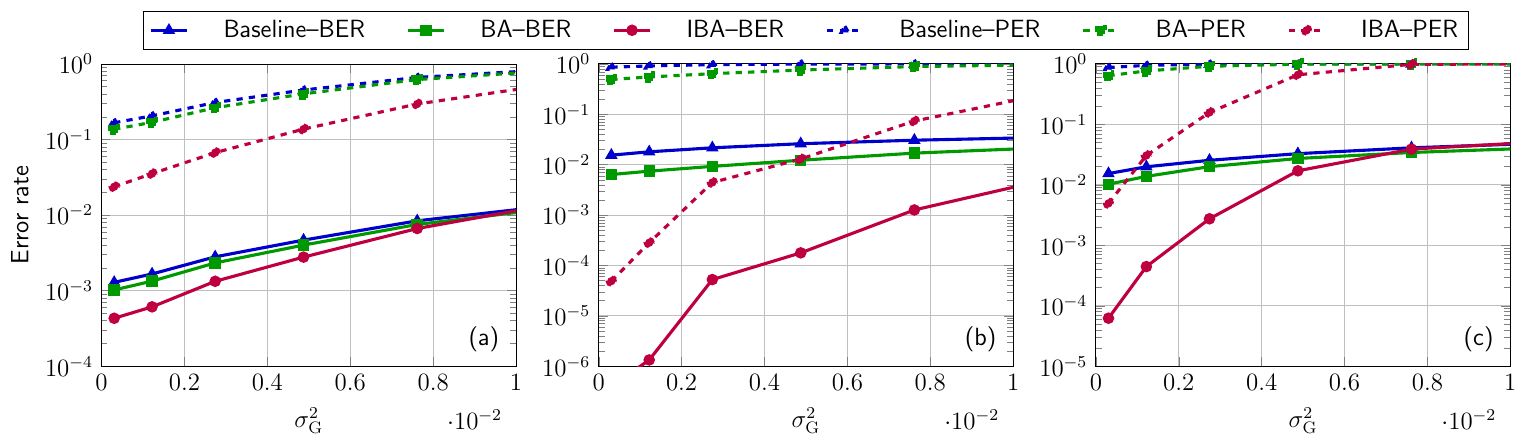}
\caption{BER and PER performance of the baseline, \ac{ba}, and \ac{iba} \ac{ldpc} decoding schemes under different $\sigma_\text{G}^{2}$ values, with fixed parameters 
$\sigma_\text{B}^{2} = 0.12$, $P_\text{GB} = 2 \cdot 10^{-4}$, 
and $P_\text{BG} = 2 \cdot 10^{-2}$. 
(a)~QPSK with $\text{SNR}= 8$~dB; 
(b)~16QAM with $\text{SNR}= 15$~dB; 
(c)~64QAM with $\text{SNR}= 20.8$~dB.}
\label{fig:ibasigmag}
\end{figure*}

\begin{figure*}
\centering
\includegraphics[scale=0.667]{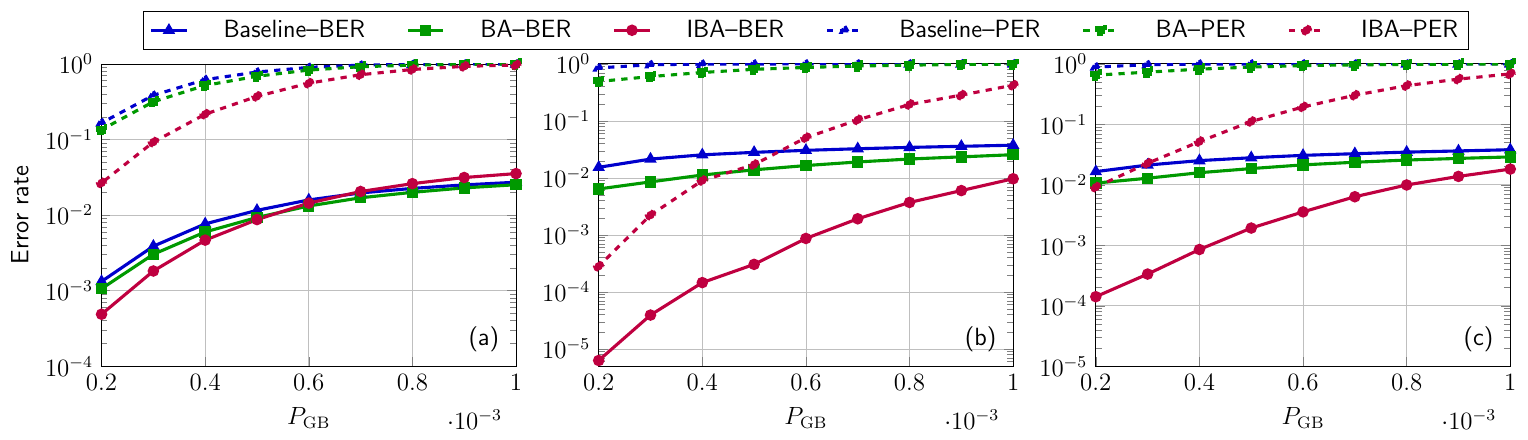}
\caption{BER and PER performance of the baseline, \ac{ba}, and \ac{iba} \ac{ldpc} decoding schemes 
under different $P_\text{GB}$ values, with fixed parameters 
$\sigma_\text{G}^{2} = 3 \cdot 10^{-4}$, $\sigma_\text{B}^{2} = 0.12$, 
and $P_\text{BG} = 2 \cdot 10^{-2}$. 
(a)~QPSK with $\text{SNR}= 8$~dB; 
(b)~16QAM with $\text{SNR}= 15$~dB; 
(c)~64QAM with $\text{SNR}= 20.8$~dB.}
\label{fig:ibapgb}
\end{figure*}

\begin{figure*}
\centering
\includegraphics[scale=0.667]{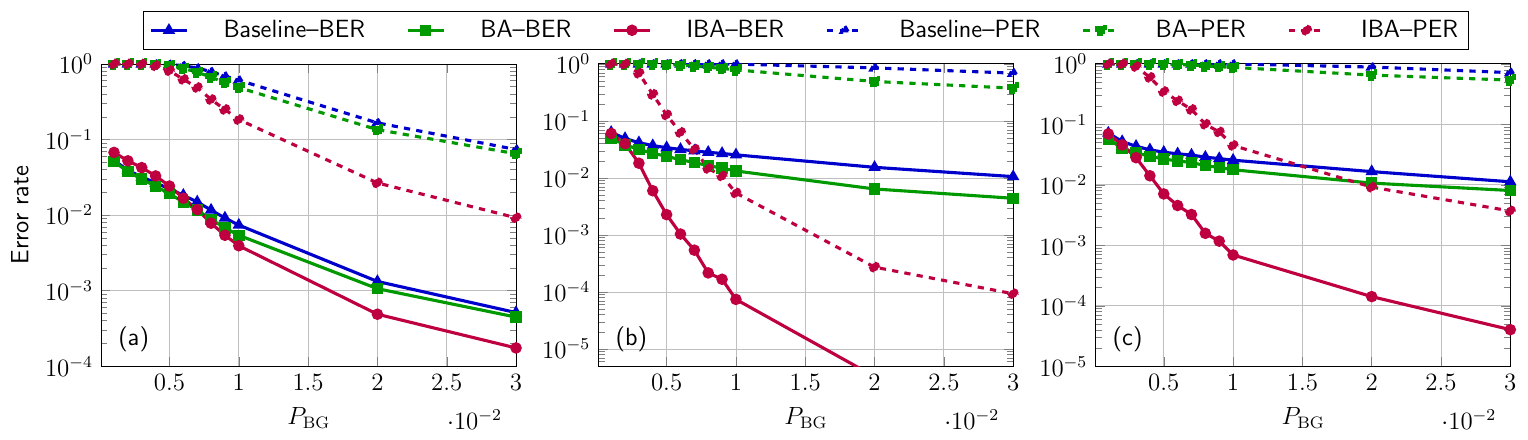}
\caption{BER and PER performance of the baseline, \ac{ba}, and \ac{iba} \ac{ldpc} decoding schemes 
under different $P_\text{BG}$ values, with fixed parameters 
$\sigma_\text{G}^{2} = 3 \cdot 10^{-4}$, $\sigma_\text{B}^{2} = 0.12$, 
and $P_\text{GB} = 2 \cdot 10^{-4}$. 
(a)~QPSK with $\text{SNR}= 8$~dB; 
(b)~16QAM with $\text{SNR}= 15$~dB; 
(c)~64QAM with $\text{SNR}= 20.8$~dB.}
\label{fig:ibapbg}
\end{figure*}

The channel information of the simulation system is shown in Fig.~\ref{fig:channel_information}(a)--(c). The phase noise generated by the Wiener-GE model is shown in Fig.~\ref{fig:channel_information}(a). 
The results show that the phase noise changes slowly most of the time, and only when the channel enters the bad state will there be obvious fluctuations.
As a result, the differential phase noise is shown as a zero-mean Gaussian distribution with different variances that depend on the channel state. The estimated channel states are illustrated in Fig.~\ref{fig:channel_information}(d)--(f). To ensure numerical stability, all three channel estimation schemes are implemented in the logarithmic domain. For practical considerations, the backtracking lengths for \ac{va} and \ac{sova} are both set to 100 in the simulation. Furthermore, the \ac{bcjr}-based estimator is implemented using a windowed \ac{bcjr} algorithm with a window size of 100. The channel state estimated by \ac{va} misses some short bursts due to its hard-decision nature. The result obtained by \ac{sova} shows noticeable fluctuations and lacks stability. In contrast, the \ac{bcjr}-based estimation closely matches the real channel state.

\subsection{\ac{ba} \ac{ldpc} decoding}
To compare the performance of the baseline and \ac{ba} schemes with three types of channel estimation, simulations were conducted for QPSK, 16QAM, and 64QAM transmission formats, as shown in Fig.~\ref{fig:snrber} and \ref{fig:snrper}.
Considering the \ac{snr} loss caused by differential coding, a fixed offset of $\delta = -3$~dB was applied in these simulations.
Both the BER and PER results consistently show that the \ac{bcjr}-based channel estimation provides the best performance among the three estimation methods. For QPSK, the four decoding schemes show similar performance, with the \ac{bcjr}-based \ac{ba} \ac{ldpc} scheme being slightly better than the others. For 16QAM, the \ac{bcjr}-based \ac{ba} scheme achieves a gain of about 0.1~dB at a BER of $4\cdot10^{-3}$ and about 0.4~dB at a PER of $10^{-2}$. For 64QAM, the \ac{bcjr}-based scheme offers a 0.7-dB \ac{snr} gain over the baseline at same BER and reaches a PER of $10^{-2}$ at around $21.5$~dB, while the baseline fails to achieve this level within the simulated range. 
According to these results, subsequent \ac{ba} and \ac{iba} \ac{ldpc} decoding schemes use the \ac{bcjr}-based channel estimation.

For complexity, compared with the baseline scheme, the BA scheme introduces a channel-estimation stage implemented by a two-state windowed BCJR algorithm, whose cost grows linearly with the sequence length.

To achieve the best overall performance of the \ac{ba} \ac{ldpc} decoding, the bias $\delta$ is optimized with three modulations as shown in
Fig.~\ref{fig:Bias1opt}. Different SNRs are used for different modulation formats to operate each system in a comparable error-rate regime.
For QPSK, the optimal bias is approximately $-3$~dB.
For 16QAM and 64QAM, the optimal bias values are around $-2$~dB.
These optimized biases $\delta$ are adopted in subsequent simulations.

\subsection{\ac{iba} \ac{ldpc} decoding}

In \ac{iba} \ac{ldpc} decoding, the bias $\delta^{\prime}$ is optimized first, and the results are shown in Fig.~\ref{fig:bias2}. The results show that QPSK is insensitive to bias, reaching its optimal performance around $\delta^{\prime} = 0$~dB. For both 16QAM and 64QAM, the performance depends strongly on $\delta’$, with the best performance achieved when $\delta^{\prime} = 5$~dB. The bias $\delta’$ is applied in computing the likelihood function used in both channel estimation and \acp{llr} calculation. During each outer iteration, the probability of each constellation point is updated. As these symbol probabilities become more accurate, the effective SNR increases.
Furthermore, for all modulation formats, performance tends to saturate after three iterations.
Table~\ref{tab:op} gives the optimized parameters used in subsequent simulations. 

The baseline, \ac{ba}, and \ac{iba} \ac{ldpc} decoding are compared in five dimensions, which are \ac{snr}, $\sigma_\text{B}^2$, $\sigma_\text{G}^2$, $P_\text{GB}$, and $P_\text{BG}$. 

\begin{table}[!t]
\caption{Optimized Parameters}
\centering
\begin{tabularx}{\linewidth}{l *{3}{>{\centering\arraybackslash}X}}
\toprule
\textbf{Parameter} & \textbf{QPSK} & \textbf{16QAM} & \textbf{64QAM} \\
\midrule
$\delta$ (Initial bias) & $-3$~dB & $-2$~dB & $-2$~dB \\
$\delta^{\prime}$ (Outer iteration bias) & $0$~dB & $5$~dB & $5$~dB \\
\text{No. of outer iterations} & 3 & 3 & 3 \\
\bottomrule
\end{tabularx}
\label{tab:op}
\end{table}

\subsubsection{SNR}
The performance of the baseline, \ac{ba}, and \ac{iba} decoding schemes 
was evaluated under different \acp{snr}, as shown in 
Fig.~\ref{fig:ibasnrber} and Fig.~\ref{fig:ibasnrper}.
For QPSK modulation, the three schemes exhibit nearly identical performance near the BER threshold of $4\cdot10^{-3}$. At a PER of $10^{-2}$, the \ac{iba} scheme gains approximately 0.1~dB over the \ac{ba} scheme and approximately 0.2~dB over the baseline scheme.
For 16QAM modulation, at the BER threshold, the \ac{iba} scheme outperforms the \ac{ba} scheme and the baseline scheme by approximately 0.4~dB and 0.7~dB, respectively. At the PER threshold, the gains are approximately 0.5~dB and 1.0~dB, respectively.
For 64QAM modulation, at the BER threshold, the \ac{iba} scheme gains approximately 0.5~dB over the \ac{ba} scheme and approximately 1.4~dB over the baseline scheme. At the PER threshold, \ac{iba} outperforms \ac{ba} by approximately 0.6~dB, and offers a gain of over 3~dB compared to the baseline scheme.

\subsubsection{$\sigma_{\textnormal{B}}^2$}
The performance comparison of the three schemes for different values of the parameter $\sigma_\text{B}^2$ is shown in Fig.~\ref{fig:ibasigmab}.
The parameter $\sigma_\text{B}^2$ represents the severity of burst errors.
We focus on the case of $\sigma_\text{B}^2 = 1$, where burst errors are quite severe. 
For QPSK, the baseline and \ac{ba} schemes exhibit almost identical performance, with BERs of approximately $6.60\cdot10^{-3}$ and $6.07\cdot10^{-3}$, and PERs of about $0.56$ and $0.52$, respectively. For the \ac{iba} scheme, the BER is about $4.32\cdot10^{-3}$ and the PER is about $0.20$. 
For 16QAM, the BER decreases from $2.12\cdot10^{-2}$ with the baseline scheme to $1.16\cdot10^{-2}$ with the \ac{ba} scheme, and further to $1.54\cdot10^{-4}$ with the \ac{iba} scheme. The PER shows a similar trend, decreasing from $0.96$ to $0.72$ and finally to $0.01$. This corresponds to a nearly two orders of magnitude reduction in both BER and PER.
For 64QAM, the BER decreases from $2.08\cdot10^{-2}$ to $1.44\cdot10^{-2}$, and finally to $5.06\cdot10^{-4}$, while the PER decreases from $0.97$ to $0.78$ and finally to $0.03$. Both BER and PER are reduced by more than one order of magnitude compared to baseline. 
Overall, the \ac{iba} decoding scheme showed the highest robustness to severe burst differential phase noise.

\subsubsection{$\sigma_{\textnormal{G}}^2$}
The performance comparison of the three schemes for different values of the parameter $\sigma_\text{G}^2$ is shown in Fig.~\ref{fig:ibasigmag}.
The parameter $\sigma_\text{G}^2$ denotes the differential phase noise variance in the good state, which is typically small since the differential phase noise in the good state is negligible. The comparison is made at $\sigma_\text{G}^2 = 5\cdot10^{-3}$, where the differential phase noise in the good state becomes significant. 
For QPSK, the BER performance of the three schemes is similar, with the baseline, \ac{ba}, and \ac{iba} schemes achieving BERs of $4.66\cdot10^{-2}$, $4.00\cdot10^{-2}$, and $2.78\cdot10^{-2}$, respectively. In terms of PER, the performance of the baseline and \ac{ba} schemes is comparable, with PERs of $0.45$ and $0.40$, while the \ac{iba} scheme achieves a lower PER of about $0.14$. 
For 16QAM, the BERs of the three schemes are $2.61\cdot10^{-2}$, $1.22\cdot10^{-2}$, and $1.78\cdot10^{-4}$, and the corresponding PERs are $0.99$, $0.76$, and $0.01$. The \ac{iba} scheme achieves reductions of about two orders of magnitude in both BER and PER compared with the baseline scheme. 
For 64QAM, the impact of severe differential phase noise is evident, resulting in similar performance among the three schemes. However, the \ac{iba} scheme is still slightly better than both the baseline and the \ac{ba} schemes.

\subsubsection{$P_{\textnormal{GB}}$}
The performance comparison for different values of the parameter $P_\text{GB}$ is shown in Fig.~\ref{fig:ibapgb}.
Recall from Sec.~\ref{sec:system_model} that a larger $P_\text{GB}$ results in a shorter average good-state duration.
The comparison focuses on when $P_\text{GB}=5\cdot10^{-4}$, which corresponds to an average good-state length of $1/P_\text{GB}=2000$ symbols.
For QPSK, the baseline, \ac{ba}, and \ac{iba} decoding schemes exhibit similar BER performance, with PERs of $0.78$, $0.69$, and $0.37$, respectively.
For 16QAM, the corresponding BERs are $2.81\cdot10^{-2}$, $1.39\cdot10^{-2}$, and $3.06\cdot10^{-4}$, while the PERs are $0.99$, $0.80$, and $0.02$.
For 64QAM, the BERs are $2.80\cdot10^{-2}$, $1.86\cdot10^{-2}$, and $1.93\cdot10^{-3}$, and the PERs are $0.99$, $0.89$, and $0.11$.
Overall, the \ac{iba} scheme slightly outperforms the baseline for QPSK, achieves about two orders of magnitude reduction for 16QAM, and about one order of magnitude improvement for 64QAM.


\subsubsection{$P_{\textnormal{BG}}$}
The performance comparison for different values of the parameter $P_\text{BG}$ is shown in Fig.~\ref{fig:ibapbg}.  Recall from Sec.~\ref{sec:system_model} that a larger $P_\text{BG}$ results in a shorter average bad-state duration, leading to shorter error bursts.
When $P_\text{BG}=1\cdot10^{-2}$, the average burst length is $1/P_\text{BG}=100$ symbols. 
Under this condition, for QPSK, the baseline, \ac{ba}, and \ac{iba} decoding schemes have similar BER performance, with PERs of $0.61$, $0.49$, and $0.18$, respectively.
For 16QAM, the BERs are about $2.54\cdot10^{-2}$, $1.33\cdot10^{-2}$, and $7.42\cdot10^{-5}$, and the PERs are approximately $0.99$, $0.79$, and $5.49\cdot10^{-3}$. 
For 64QAM, the BERs are approximately $2.56\cdot10^{-2}$, $1.78\cdot10^{-2}$, and $6.92\cdot10^{-4}$, with corresponding PERs of approximately $1.00$, $0.88$, and $0.04$.
In summary, the \ac{iba} scheme slightly outperforms the baseline for QPSK, achieves more than two orders of magnitude reduction for 16QAM, and provides about one order of magnitude improvement for 64QAM.

\section{Conclusions}
\label{sec:conclusion}
This paper studied channel estimation and LDPC decoding under bursty differential phase noise and proposes the BA and IBA schemes. Results show that the BCJR-based estimator provides the most reliable channel state, making the BA scheme outperform the baseline scheme; while information exchange during iterations enables the IBA scheme to have the highest robustness under bursty phase noise conditions. The results show the \ac{iba} can reduce both \ac{ber} and \ac{per} by up to two orders of magnitude in systems with severe bursty phase noise. 

Since this system relies on differential coding, which naturally introduces a certain \ac{snr} penalty, the proposed schemes are most beneficial in scenarios where \ac{per} is the primary concern, the performance is limited by severe bursty phase noise, and where the \ac{snr} is not as severe. For future work, an \ac{snr}-sensitive system may be investigated by replacing differential coding with a phase recovery algorithm. Additionally, the core concept of the \ac{ba} and \ac{iba} schemes can be extended to other bursty impairments, such as polarization-dependent bursts, EEPN, and MPI.
\normalem
\bibliographystyle{ieeetr}
\bibliography{reference.bib}

\vfill

\end{document}